\documentclass[pra,twocolumn,showpacs,superscriptaddress,groupedaddress,showpacs,shokeys]{revtex4} % for checking your page length
\usepackage{graphicx}
\usepackage{epstopdf}
\usepackage{amsmath,amsthm,amssymb}
\usepackage{color}
\usepackage{amsfonts}
\usepackage{subfigure}
\usepackage{bm}
\usepackage{cleveref}

\draft % marks overfull lines with a black rule on the right

\begin{document}

%\preprint{}

\title{Pulsed high magnetic field measurement via a Rubidium vapor sensor} %Title of paper



%\thanks{}
\author{S.George}
\affiliation{Laboratoire
National des Champs Magn\'etiques Intenses (UPR 3228,
CNRS-UPS-UGA-INSA), F-31400 Toulouse Cedex, France}
\author{N. Bruyant}
\affiliation{Laboratoire
National des Champs Magn\'etiques Intenses (UPR 3228,
CNRS-UPS-UGA-INSA), F-31400 Toulouse Cedex, France}
\author{J. B\'eard}
\affiliation{Laboratoire
National des Champs Magn\'etiques Intenses (UPR 3228,
CNRS-UPS-UGA-INSA), F-31400 Toulouse Cedex, France}
\author{S. Scotto}
\affiliation{Laboratoire
National des Champs Magn\'etiques Intenses (UPR 3228,
CNRS-UPS-UGA-INSA), F-31400 Toulouse Cedex, France}
\affiliation{Dipartimento di Fisica ``E. Fermi'', Universit\`a di Pisa, Largo B. Pontecorvo 3, 56127 Pisa, Italy}
\author{E. Arimondo}
\affiliation{Laboratoire
National des Champs Magn\'etiques Intenses (UPR 3228,
CNRS-UPS-UGA-INSA), F-31400 Toulouse Cedex, France}
\affiliation{Dipartimento di Fisica ``E. Fermi'', Universit\`a di Pisa, Largo B. Pontecorvo 3, 56127 Pisa, Italy}
\affiliation{INO-CNR, Via G. Moruzzi 1, 56124 Pisa, Italy}
\author{R. Battesti}
\affiliation{Laboratoire
National des Champs Magn\'etiques Intenses (UPR 3228,
CNRS-UPS-UGA-INSA), F-31400 Toulouse Cedex, France}
\author{D. Ciampini}
\affiliation{Dipartimento di Fisica ``E. Fermi'', Universit\`a di Pisa, Largo B. Pontecorvo 3, 56127 Pisa, Italy}
\affiliation{INO-CNR, Via G. Moruzzi 1, 56124 Pisa, Italy}
\affiliation{CNISM, UdR Dipartimento di Fisica ``E. Fermi'', Universit\`a di Pisa, Largo B. Pontecorvo 3, 56127 Pisa, Italy}
\author{C. Rizzo\footnote{carlo.rizzo@lncmi.cnrs.fr}}
\affiliation{Laboratoire
National des Champs Magn\'etiques Intenses (UPR 3228,
CNRS-UPS-UGA-INSA), F-31400 Toulouse Cedex, France}


\date{\today}

\begin{abstract}
We present a new technique to measure pulsed magnetic fields based on the use of Rubidium in gas phase as a metrological standard. We have therefore developed an instrument based on laser inducing transitions at about 780~nm (D2 line) in a Rubidium gas contained in a mini-cell of 3~mm~x~3~mm cross section. To be able to insert such a cell in a standard high field pulsed magnet we have realized a fibred probe kept at a fixed temperature. Transition frequencies for both the $\pi$ (light polarization parallel to the magnetic field) and $\sigma$ (light polarization perpendicular to the magnetic field) configurations are measured by a commercial wavemeter. One innovation of our sensor is that in addition of monitoring the light transmitted by the Rb cell, which is usual, we also monitor the fluorescence emission of the gas sample from a very small volume with the advantage of reducing the impact of the field inhomogeneity on the field measurement. Our sensor has been tested up to about 58~T.
\end{abstract}


\maketitle %\maketitle must follow title, authors, abstract and \pacs

% References should be done using the \cite and \label commands
\section{Introduction}

While several methods to measure precisely a magnetic field exist \cite{Symonds1955}, nowadays a very accurate measurement of magnetic field is performed via the nuclear magnetic resonance (NMR) of hydrogen in a water molecule. Devices based on other techniques are calibrated with respect to NMR. According to the NMR technique, the value of the magnetic field $B$ experienced by the hydrogen nucleus, i.e. the proton, is derived by the frequency $\nu_{NMR}$ of the microwave inducing at resonant spin flip of the proton
\begin{equation}\label{NMR}
   \nu_{RMN} = \frac{\gamma^{'}_p}{2\pi} B
\end{equation}
where $\gamma^{'}_p$  is the gyromagnetic factor of the proton in water. The measurement of $\gamma^{'}_p$ with respect to the electron magnetic moment has been first performed in water at 34.7~$^{\circ}$C by Phillips, Cooke and Kleppner \cite{Phillips1977} in 1977 at a field of about 0.35~T. The recommended value of $\gamma^{'}_p$ given in \cite{CODATA} is  ${\gamma^{'}/(2\pi)} =  42.57638507(53)$~MHz.T$^{-1}$ (Water, sphere, 25~$^{\circ}$C). In commercial devices, for fields higher than 10~T the H$_2$O molecule is replaced by the D$_2$O molecule, which has a lower gyromagnetic factor than hydrogen, to keep the spin flip resonance frequency lower than 500~MHz (see e.g.\cite{MetrolabPT2026}). This kind of apparatus is designed to measure continuous fields greater than 0.2~T over a volume of a few mm$^3$ with a precision better than 1~ppm and an accuracy of 5~ppm.

In the case of pulsed fields, \textit{i.e.} fields varying on a timescale shorter than a second, pulsed NMR techniques have been developed recently to be used as a probe of matter properties (see \cite{Stork2013} and refs within), but not yet for metrological purposes.

In this paper, we present a new technique to measure pulsed magnetic fields based on the optical transitions of Rubidium in gas phase as a metrological standard.
Optical magnetometry based on Rubidium vapor is already used to measure very low magnetic fields because of the precise Rubidium atomic parameters for its ground state~\cite{Budkerlibro}. Our goal is to extend to high magnetic fields the measurement capability of Rubidium gas by monitoring the optical transition frequencies  between ground state and first excited states. As in the case of NMR, at large applied magnetic fields the transition frequency  between well chosen quantum levels  depends  linearly on the applied magnetic field. We have therefore developed an instrument based on a narrow band and stable laser inducing the D2 transition for Rubidium gas contained in a mini-cell within a volume of 0.13~mm$^3$. To be able to insert such a cell in a standard high field pulsed magnet we have realized a fibred probe kept at a fixed temperature. Transition frequency both in the $\pi$ (light polarization parallel to the magnetic field) and $\sigma$ (light polarization perpendicular to the magnetic field) configurations are measured by a commercial wavemeter.

The design of our sensor follows the pioneering work initiated at NIST-Boulder around 2004 in order to develop miniaturized atomic clocks \cite{NIST} as reviewed in detail by Budker \textit{et al}~\cite{Budkerlibro}. These microfabricated magnetometers have been used to detect fields of the order of pT produced by the human body and in the domain of low field NMR for remote imaging or chemical species investigation \cite{Budkerlibro}. A particularity of our sensor is that in addition of monitoring the light transmitted by the Rb cell, which is usual, we also monitor the fluorescence emission of the gas sample with the advantage to reduce the impact of the field inhomogeneity on the field measurement. Our sensor was tested up to about 58~T which represents the highest field value to which a gas sample has been exposed using non destructive field generation. Actually, as far as we know, the only other attempt similar to ours dates back to 1971~\cite{King1971} when a field pulsed up to 33~T in less than a millisecond has been measured by monitoring the mercury line at 253.7~nm. Averaging several hundred measurements an uncertainty on the field value of about 0.04~$\%$ has been obtained, which, as far as we understand, translates into an uncertainty of less than 0.1~$\%$ per pulse. This uncertainty was essentially limited by the uncertainty on the transition frequency measurement.

In the case of magnetic fields obtained in a destructive way, measurements using spectroscopic techniques date back to 1966 when a field of about 500~T obtained by explosive flux compression has been measured observing Sodium and Indium lines \cite{Garn1966}.  More recently, fields in excess of 200~T has been measured by observing the splitting of sodium doublet around 589 nm in the case of magnetically imploded targets for inertial confinement fusion \cite{Gomez2014}. Spectroscopy of Sodium atoms has also been used in the case of fields produced by exploding wires to measure fields around 50~T in the eighties \cite{Hori:1982} and 20~T very recently \cite{Banasek2016}. Those investigations were limited by the collisional and thermal broadening of the absorption lines due to the explosion process. This is not the case for the present investigation.

In section 2 we present our method to perform a full-optical magnetic field measurement based on alkali atoms. The following section 3 describe the pulsed magnetic field coils that have been used in the present work. The section 4 is devoted to the probe and the sensor design that we explain in details. In section 5 we explain how we have calibrated a pick-up coil that we have used to monitor the magnetic field pulse.  The experimental set-up is described in section 6. Finally the section 7 presents the results we obtained in terms of spectroscopy signals up to a field of about 58~T and and in terms of comparison between field values given by our Rb sensor and the standard pick-up coil. A final section concludes our paper including perspectives and applications of the present work.

\section{Method}
\subsection{Full-optical magnetic field measurement}

Our method can be easily presented for the optical second resonance line $5^2S_{1/1} \to 5^2P_{3/2}$ of an idealized alkali atom without nuclear spin, and therefore without nuclear Zeeman shift and without the electron-nucleus hyperfine coupling~\cite{Arimondo:1977,ArimondoAmop}. In the following, we will consider only the linear Zeeman effect assuming that the second order Zeeman effect~\cite{ArimondoAmop} due to diamagnetism can be neglected at the field strengths of interest.

For an applied magnetic field $B$, the $E_g$ eigenvalues for the $[J_g=1/2,m_g=\pm 1/2]$ ground eigenstates are determined by the Zeeman magnetic coupling and given by
\begin{equation}
\frac{E_g(J_g,m_g)}{h}=\mu_Bg_{5S}m_gB,
\label{GroundEnergies}
\end{equation}
where $\mu_B$ is the Bohr magneton in MHz/T ($\mu_B = 13996.245042(86)$~MHz/T in~\cite{CODATA}) and $g_{5S}\approx 2$ the electron ground g-factor.  The ground state splitting for  fields of a few tenth of tesla is of the order few hundreds of GHz.
The frequency determination of that splitting in the Rubidium ground state measures directly the magnetic field.\\
\indent Also the $E_e$ eigenvalues for the $|J_e=3/2,m_e= (\pm3/2,\pm1/2)\rangle$ excited eigenstates are determined by the Zeeman magnetic coupling and given by
\begin{equation}
\frac{E_e(J_g,m_g)}{h}=\mu_Bg_{5P}m_eB,
\label{ExcitedEnergies}
\end{equation}
where $g_{5P}$ is the electron Land\'e g-factor for the $5^2P_{3/2}$ state. Because $g_{5P}\approx 4/3$, as discussed in the following, the excited state Zeeman splittings are similar to those in the ground state.\\
\indent The optical transition between the $|J_g,m_g\rangle \to |J_e,m_e\rangle$ states experiences a magnetic field Zeeman frequency shift $\Delta\nu_{Z}$ given by
\begin{equation}
\Delta\nu_{Z}=\mu_B (m_eg_{5P}-m_gg_{5S})B,
\label{DeltanuZeeman}
\end{equation}
and $\nu$ optical transition frequency is given by
\begin{equation}
\nu^{Rb}=\nu_0^{Rb}+\Delta\nu_{Z},
\label{OpticalZeeman}
\end{equation}
 where $\nu_0^{Rb}$ represents the center of gravity for the D2 absorptions in either $^{85}$Rb or $^{87}$Rb reported in~\cite{Steck85:2001} and \cite{Steck87:2001}. For optical $\sigma^{\pm}$ or $\pi$ polarized transitions given by $m_e=m_g\pm1$ or $m_e=m_g$, respectively,  the Zeeman shift appearing in this equation is comparable to those quoted above for the ground state. The inversion of Eq.~\eqref{DeltanuZeeman} allows to derive the magnetic field from a measurement of $\Delta\nu_{Z}$.  The high sensitivity associated to the measurement of optical absorption processes and the resulting fluorescence emission leads to an efficient application of the optical detection. \\
%\indent Neglecting the uncertainty on the $\mu_B$, the magnetic field uncertainty $u(B)$ can be written as
%\begin{equation}
%\frac{u(B)}{B}= \sqrt{\left(\frac{ u(\Delta\nu_{Z})}{\Delta\nu_{Z}}\right)^2+\left(\frac{m_eu(g_{5P})+m_gu(g_{5S})}{m_eg_{5P}-m_gg_{5S}}\right)^2},
%\label{Err}
%\end{equation}
%where $u(\Delta\nu_{Z})$ is the uncertainty in the Zeeman shift measurement,  $u(g_{5P})$ and $u(g_{5S})$ represent the indetermination for the $g$-factors of the lower and upper state, respectively. The very high precision associated to the measurement of optical frequencies allows to reach a 0.1~ppm accuracy if the Rubidium atomic constants are known at the same level.\\
\indent The experimental method is quite simple. Laser light kept at a fixed frequency excites Rubidium gas contained within a cell inserted in a magnet delivering a pulsed magnetic field.  The transition of interest between two selected Rubidium states is monitored all along the pulse evolution in order to determine the time when the magnetic field satisfies Eq. \eqref{DeltanuZeeman}.  The temporal form of the pulse may also be monitored using a pick-up coil. The value of the magnetic field given by the pick up signal is therefore calibrated through the signal produced by the Rubidium gas cell.\\

\subsection{Alkali atoms in a magnetic field}

The Rubidium ground state eigenenergies in a magnetic field do not satisfy the simple relation of Eq.~\eqref{nu}. In fact the two stable Rubidium isotopes, $^{85}$Rb and $^{87}$Rb, have a nuclear moment, $I=5/2$ and $I=3/2$, respectively, characterized by the nuclear Lande g-factor $g_I$ (assumed  negative in the following as in~\cite{Steck85:2001,Steck87:2001,ArimondoAmop}). The eigenenergy contributions by the electron-nucleus hyperfine coupling and the Zeeman nuclear energy lead to complex  functional dependencies on the magnetic field. The magnetic response is characterized by the ratio between the magnetic interactions of electron and nucleus with the magnetic field, and the hyperfine electron-nucleus coupling. For the alkali $J_g=1/2$ ground state, analytical expressions of the eigenenergies are given by the Breit-Rabi formula~\cite{BreitRabi:1931,Foot:2012}. For the excited state, no analytical formula exists and a numerical approach is necessary to diagonalize the Hamiltonian to obtain the eigenenergies at a given  magnetic field. \\
\indent Our magnetic field measurement is based on the existence of two eigenstates,  ground and excited, whose energy dependence on the magnetic field is always linear. Those eigenstates will be denoted as extreme in the following. Within the magnetic field regime here explored both  ground and excited states are characterized by the electronic angular momentum  $\mathbf{J}$  and its projection $m_J$  along the magnetic field axis, combined with nuclear moment  $\mathbf{I}$ and its projection $m_I$. The extreme eigenstates correspond to the highest values of all these quantum numbers. The $|J_g=1/2,m_g=1/2;I,m_I=I\rangle$ ground   has the following energy derived from the Breit-Rabi formula:
\begin{equation}
\frac{E_g^+}{h} =\mu_B\left(\frac{g_{5S}}{2}+g_II\right)B+\frac{1}{2}A_gI
\label{EstrGr}
\end{equation}
where $A_g$ is the dipolar hyperfine coupling of the ground state.\\
\indent Also for the excited state $5P_{3/2}$,  in the hyperfine Paschen-Back regime the eigenstates are characterized by those quantum numbers.  The $|J_e=3/2,m_e=3/2;I,m_I=I\rangle$ state   has the following energy derived from the Paschen-Back formula:
\begin{equation}
\frac{E_e^+}{h} =\mu_B\left(\frac{3g_{5P}}{2} + g_II\right)B +\frac{3}{2}A_eI+2B_e,
\label{EstrExc}
\end{equation}
where $A_e$ and $B_e$ are  the dipolar and quadrupolar hyperfine couplings of the  $5P_{3/2}$ state.\\
\indent  An additional regime, denoted as the fine Paschen-Back one,  is reached when the electron Zeeman energy is larger than the fine structure splitting between the $5P_{3/2}$ and $5P_{1/2}$ excited states. In that regime a linear dependence on $B$ applies to all the eigenenergies. For the Rubidium case this last regime cannot be reached for magnetic field presently available in earth Laboratories. Notice that owing to the smaller fine structure splitting, that regime was explored for sodium atoms in all the high field experiments with exploding wires~\cite{Garn1966,Hori:1982,Gomez2014,Banasek2016}.\\
\indent Combining together Eqs. \eqref{EstrGr} and  \eqref{EstrExc}, the frequency of the Zeeman shifted $\sigma^+$ optical transition linking the Rb linear dependent states is given by
\begin{equation}
\Delta \nu_{Z}=\mu_BB\frac{3g_{5P}-g_{5S}}{2}+ \frac{I}{2}(3A_e-A_g)+2B_e,
\label{nu}
\end{equation}
That represents the generalization of Eq.~\eqref{DeltanuZeeman} to an alkali atom as Rubidium.
Inverting this relation and making use of Eq.~\eqref{OpticalZeeman} for the Rubidium case, we obtain the following relation determining the magnetic field for a measured $\nu^{Rb}$ optical frequency:
\begin{equation}
B=\frac{2(\nu^{Rb}-\nu_0^{Rb})}{\mu_B(3g_{5P-}g_{5S})}- \frac{(3A_e-A_g)I}{\mu_B(3g_{5P-}g_{5S})}-\frac{4B_e}{\mu_B(3g_{5P-}g_{5S})}.
\label{field}
\end{equation}

\subsection{Rubidium atomic constants}

Our optical determination of the magnetic field  is based on the D2 transition between the ground state 5$^2$S$_{1/2}$ and the excited state 5$^3$P$_{3/2}$ of the rubidium isotopes, whose $\nu_0^{Rb}$ frequency separation  at zero magnetic field is known with a $1\times10^{-11}$ precision~\cite{Steck85:2001,Steck87:2001}. We therefore need to know accurately ground and excited atomic constants.
The ground state hyperfine splitting is $A_{5S} = 3.417341305452145(45)$~GHz for $^{87}$Rb as measured by Bize {\it et al.}~\cite{Bize1999}, while for  $^{85}$Rb the $A_{5S} = 1.011 910 813 0(20)$~GHz value was reported by \cite{Arimondo:1977,Steck85:2001}. The ground state Land\'e g-factor was precisely measured for $^{87}$Rb by Tiedemann and Robinson in 1977~\cite{Tiedemann1977} with respect to $g_e$, the free electron g-factor. The ratio reported ratio was $R_{Rb} = g_{5S}/g_e = 1.000005876(13)$, measured at 5~mT. Since isotopic effects on Rubidium g-factor were found to be less than 1~ppb \cite{Arimondo:1977}, this ratio also apply to $^{85}$Rb within the errors. Making use of the $g_e$ value given in~\cite{CODATA}  we obtain
\begin{equation}
g_{5S} = R_{Rb} g_e = 2.002331070(26)
\label{6}
\end{equation}
\indent For the 5$^3$P$_{3/2}$ excited state, the $^{87}$Rb dipolar and quadrupolar hyperfine constants carefully measured by Ye \textit{et al}.~\cite{YeHall1996} and reported in~\cite{Steck87:2001} are $A_e=84.7185(20)$~MHz and $B_e=12.4965(37)$~MHz, while for the $^{85}$Rb ones those reported in~\cite{Arimondo:1977,Steck85:2001} remain the most recent determinations. A large indetermination is associated to the Land\'e $g_{5P}$-factors with respect to the $g_{5S}$ one.  The data of ~\cite{Arimondo:1977} point out that for all the alkali atoms, within the reported experimental errors, the $g$-factor of the first excited $P$ state is  $\approx 1.33411$, the value predicted by the Russel-Saunders coupling.  For the $5P_{3/2}$ state in $^{87}$Rb ref.~\cite{Arimondo:1977} reported 1.3362(13) as a weighted average of all the measurements available at that time. No new measurement is available. That value is largely determined by fitting the level crossing measurements by Belin and Svanberg~\cite{BelinSvanberg1971}, and deriving at the same time also the dipolar and quadrupolar hyperfine constants of that state. We have reanalysed those level-crossing measurements  by fixing the hyperfine constants  to the very precise values of ref.~\cite{YeHall1996} and the $^{87}$Rb nuclear magnetic moment to the value of ref.~\cite{WhiteRobinson1973}.  A new $g_{5P3/2}=1.3341(2)$ value was obtained, in agreement with the Russel-Saunders prediction.  Following ref. \cite{Flambaum}, we evaluate the QED and relativistic corrections  at the level of 10$^{-4}$-10$^{-5}$. Therefore  we will use that value in our analysis.  \\
\indent At a magnetic field of 50~T  the predicted Zeeman shift is around 700~GHz.   Inserting the g-factor uncertainties and a planned 10~MHz accuracy (equivalent to 1/50 of the Doppler width) in the determination of the resonance frequency, we estimate that our determination leads to an accuracy of about 10$^{-4}$ for a 50~T magnetic field.\\
\indent This accuracy is orders of magnitude better of the one quoted in previous high magnetic field studies of atoms~\cite{Garn1966, Hori:1982, Gomez2014, Banasek2016}. None of these previous investigations at very high magnetic fields has included the diamagnetism in their analysis. However, the presence of a diamagnetic correction for  experiments to be performed at magnetic fields higher than those presented in this work should be taken into account. This could also be true if a large accuracy is aimed. Diamagnetic constants have only been measured for high excited states of alkalis ~\cite{ArimondoAmop} and no theoretical prediction exists yet for the ground state or excited $P$ states.

\subsection{Magnetic field measurement uncertainty}
In the following, we apply the recommendations of the Joint Committee for Guides in Metrology (JCGM)~\cite{JCGM2012}, and all uncertainties are given for a coverage factor of $1$.

Neglecting the uncertainty on $\mu_B$,  $A_e$, $A_g$ and $B_e$, following Eq.~\eqref{field} the magnetic field uncertainty $u(B)$ is determined by two separated contributions produced by the $u(\Delta\nu^{Rb})$ uncertainty in laser frequency difference between the zero field transition and the B field one,  and the $u(g_{5P})$ and $u(g_{5S})$  uncertainties for the $g$-factors of the lower and upper state. These uncertainties have different sources and lead to A type and B type uncertainties. Supposing the frequency reading described by a Gaussian with $u(\Delta\nu^{Rb})$ variance we write for the A type contribution:
\begin{equation}
\frac{u(B)_{A}}{B}= \frac{ u(\Delta\nu^{Rb})}{\Delta\nu^{Rb}}
\label{Errrand}
\end{equation}
The $g$-factors are instead affected by B type uncertainties $u(g_{5P})$ and $u(g_{5S})$, with $u(g_{5S})$ being negligible with respect to $u(g_{5P})$, and produce the following B type contribution to the field uncertainty:
\begin{equation}
\frac{u(B)_{B}}{B}= \frac{3u(g_{5P})}{\mu_B(3g_{5P-}g_{5S})},
\label{Errsyst}
\end{equation}
Within our analysis we will add in quadrature the above contributions. The very high precision associated to the measurement of optical frequencies could allow to reach a 0.1~ppm uncertainty if the Rubidium atomic constants are all known at the same level, which is not the case at the moment as shown in the previous subsection.

\section{Pulsed magnetic field coil}

\begin{figure}[ht]
\centering
 \includegraphics[width=8cm]{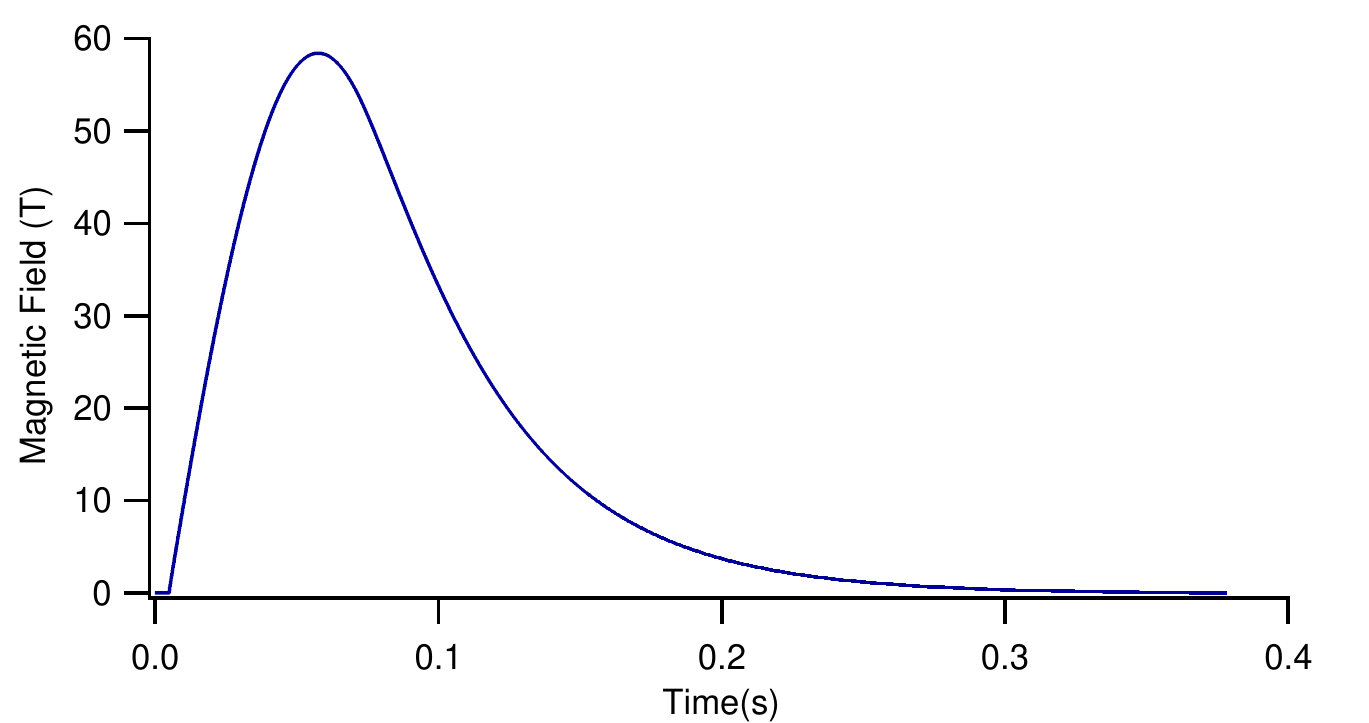}

  \caption{Typical magnetic field pulse corresponding to a maximum field around 59~T.}\label{B}
\end{figure}

The pulsed magnetic field coil used in this experiment is a LNCMI standard 60 T coil \cite{Debray2013}. It consists in 24 layers of 40 turns each composed of 9.6~mm$^2$ hard copper wire of rectangular cross section reinforced with Toyobo Zylon fibers using the distributed reinforcement technique \cite{vanBockstal1991}. The winding outer diameter is 270~mm and the length is 160~mm. This magnet has a 28~mm free bore diameter to perform experiments. To facilitate the heat dissipation the  pulsed magnet is immersed into liquid nitrogen. To maintain the Rubidium cell at room temperature the probe is placed in a double-walled stainless steel cryostat inserted in the magnet bore. Due to the space occupied by the insulation walls and the intermediate vacuum, the bore diameter in the magnetic field is 21~mm. The magnet is connected to a capacitor bank and needs 10~kA representing 3~MJ of magnetic energy to generate 60~T. The rise time of magnetic field is about 55~ms and the time between two consecutive pulses at maximum field, necessary for the coil to cool down is one hour thanks to an annular cooling channel inserted directly in the winding \cite{Frings2008}.
Figure \ref{B} shows a typical magnetic field pulse corresponding to a maximum field of about 59~T. The magnetic field homogeneity at the center of the field region is estimated to be better than 100~ppm on 1~mm$^2$.

\label{Bfield}

\section{Probe and sensor design}

\begin{figure}[ht]
 \centering
\includegraphics[width=8cm]{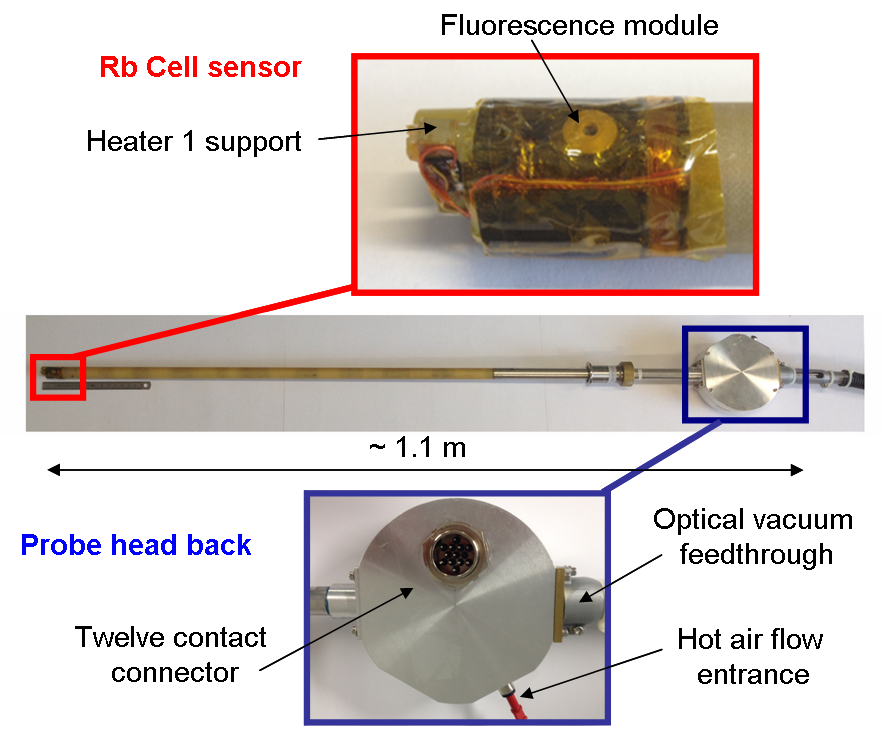}
\caption{In the center, view of the whole probe. In an expanded scale the Rb cell sensor on the top, and the probe head on the bottom}
\label{Probe}
\end{figure}

The probe head is composed by Rb cell sensor located at the end of a long pipe terminating on a chamber hosting all the electrical and optical connections, as represented into the central part of  Fig.~\ref{Probe}. The long metallic pipe is required in order to center the cell within the magnet and have the connectors outside the magnet. The sensor overall dimensions are 40~mm length and 19~mm diameter.
 The probe head contains three home-made optical vacuum-type feedthrough without fiber discontinuity and all the electrical connections, thanks to a twelve-contact connector.

\begin{figure}[ht]
\center
\includegraphics[width=8cm]{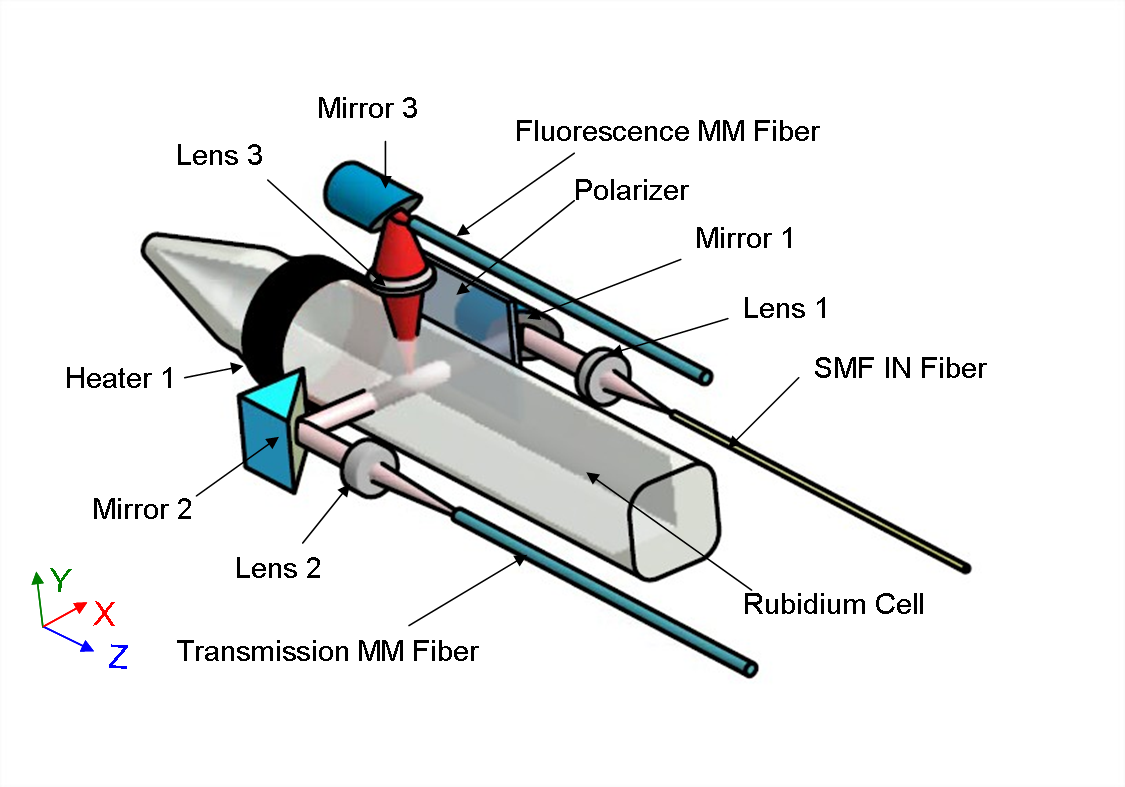}
\includegraphics[width=8cm]{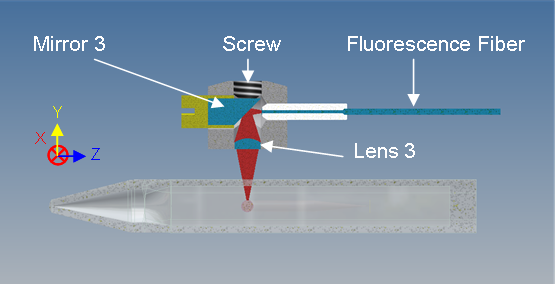}
 \caption{On the top optical scheme of our sensor. On the bottom a schematic side view of the fluorescence module optics.}
\label{Schemes}
\end{figure}

\subsection{Sensor}

\begin{figure}[ht]
\hspace{-1.5cm}
\center  % Requires \usepackage{graphicx}
 \includegraphics[width=7cm]{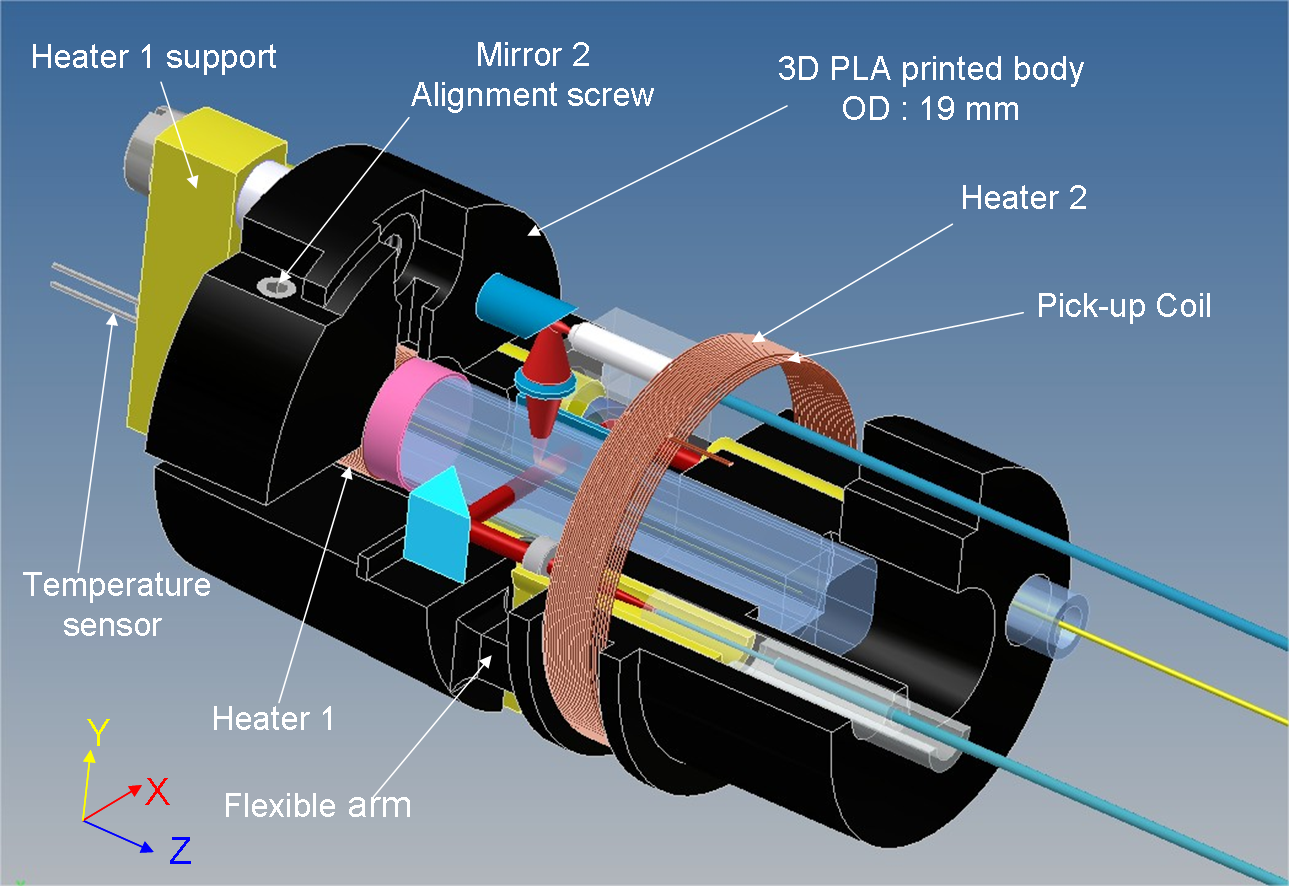}
  \includegraphics[width=6cm]{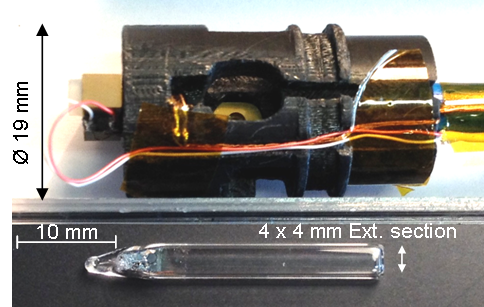}
\caption{On the top sketch of the PLA structure hosting the sensor optics. On the bottom photo of the PLA structure and of the Rb cell.}
\label{Head}
\end{figure}

The central part of the sensor is a Rubidium cell of 3~mm $\times$ 3~mm internal cross section and 30~mm length, as schematized in Fig.~\ref{Schemes}. The cell is filled with natural Rubidium therefore containing both $^{85}$Rb and $^{87}$Rb isotopes. Laser light arrives into the cell via a single mode optical fiber (SMF IN Fiber) passing trough a plan-convex lens (Lens 1)  of 2~mm diameter and focal length of 4~mm to be collimated into the vapor region after reflection on an aluminum coated 45$^{\circ}$ rod mirror (Mirror 1) of 2~mm diameter. Before entering into the cell the light is polarized at 45$^{\circ}$ with respect to the magnetic field direction to be able to induce both $\pi$ and $\sigma$ transitions by a 5~mm $\times$ 4~mm Nano-Particle Glass Polarizer slab (Polarizer) of 0.26~mm thickness. Light passing through the gas after reflection on 3~mm diameter aluminum coated N-BK7 right angle prism mirror (Mirror 2) and after being focused by an aspheric lens (Lens 2) of 5~mm focal length is collected by a 0.39 numerical aperture, 0.2~mm core multimode optical fiber (Transmission MM Fiber). This allows us to monitor the transmission of the Rubidium gas. Mirror 1 mount is coupled to an external precision mount allowing $Z$ rotation and $Z$ translation. Mirror 2 is glued on a flexible arm allowing $X$ and $Y$ rotation. Such flexible arm is also visible within the overall sketch of the structure hosting the sensor reported in Fig.~\ref{Head} top.\\
\indent At resonance Rubidium atoms absorb photons by changing their internal state from the ground level to the excited one. This excitation energy is then released as fluorescence. A particularity of our microfabricated magnetometer is to collect part of this fluorescence in a solid angle of about $4\pi/50$~sr.  The fluorescence detected in this way and generated from an atomic volume of about 0.13~mm$^3$, around one million atoms, is focused by a plan-convex lens (Lens 3) of 2.5~mm diameter and focal length of 2~mm, which is situated at 3.1~mm from the end of the optical fiber, 2.3~mm from the Rb cell and 4.3~mm from its center as sketched in the bottom of Fig.~\ref{Schemes}. Fluorescence light is then collected by a 0.39 numerical aperture, 0.2~mm core multimode optical fiber (Fluorescence MM Fiber) after being reflected by an aluminum coated 45$^{\circ}$ rod lens (Mirror 3) of 2~mm diameter. Since we will use mainly the fluorescence signal to determine the magnetic field, the volume of vapor being at the origin of the fluorescence signal gives also the spatial sensitivity of our system. The fluorescence module, sketched in the bottom part of Fig.~\ref{Schemes}, constitutes a separate part that is aligned before operation using light propagating in the opposite direction to have a focus at about 4~mm from lens 3. Thanks to the screw threading shown in the bottom of Fig. \ref{Schemes} an external precision mount is coupled to the fluorescence module allowing a proper positioning  of the module along y and z axis. Once transmission and fluorescence aligned, all the optics is glued to the PLA structure and the external precision mounts are removed. The flexible arm on which Mirror 2 is mounted allowing its rotation is also glued. All the above optical elements are hosted on a structure built thanks to a PolyLactic Acid (PLA) fused filament deposition by a 3-D printer as shown in Fig. \ref{Head} top and bottom.

\subsection{Sensor temperature control}

During operation, the sensor is placed in a cryostat inserted in the pulsed field magnet, as explained before. This kind of magnets is cooled with liquid nitrogen, even if the cryostat have a good level of thermic isolation, at the position of the sensor the temperature can be several degrees under 0~$^{\circ}$C.
In contact with the cell a heating system (Heater 1 in the top of both Fig.~\ref{Schemes} and Fig.~\ref{Probe}) is placed to control the Rubidium temperature. During standard operation the power consumption of this heater is around 200~mW.

A second heater (Heater 2 in Fig.~\ref{Head}) driven by about 100~mW power, surrounds the whole sensor and it is used in parallel with the first one to stabilize the gas temperature. It also participates to the effort to keep Rubidium temperature around 30~$^{\circ}$C, as measured by the temperature sensor located in contact with the gas cell.
All our heaters mounted are fabricated by winding 0.1~mm diameter wires of manganin alloy.
A third heater consisting of a hot air flow entrance injected through the probe, see bottom part of Fig.~\ref{Probe}, is added to the head to increase the total heating power of the sensor.

\section{Pick-up coil calibration}

Pulsed magnetic fields are usually monitored with in situ pick-up coils. One of them, consisting in 21 turns of copper wire is therefore also hosted by the PLA structure (see Fig.~\ref{Head}). It is winded in an insulating mandrel designed so that the pick up coil gives a signal that is proportional to the time variation of the magnetic field flux through a surface that is perpendicular to the field direction. Its frequency response corresponds to a bandwidth larger than 500~kHz.  For practical reasons the pick-up coil is situated at a distance of 7.5~mm from the volume of gas from which the fluorescence is originated.
To provide the time profile of the magnetic pulse one has to integrate the pick-up signal. Once calibrated, a pick-up coil can be also used as a magnetometer. To this purpose, the evaluation of the total area of this pick-up coil is realized by inserting it in a magnetic field provided by a calibration solenoid whose geometrical properties are summarized in Table \ref{TSol}.

 \begin{table}
\caption{Geometrical characteristics of the calibration solenoid. }
\label{TSol}
\begin{tabular}{cc}
\hline
  Useful inner diameter & 28 mm \\
  Number of  layers & 2 \\
  Number of turns per layer & 1463 \\
  Wire diameter including insulation & 0.34 mm \\
  Winding length  & 506.00(17) mm  \\
  Number of turns per meter & 5783(2) m$^{-1}$ \\
  First layer inner diameter  & 30.10(2) mm  \\
  Second layer inner diameter& 30.8(1) mm  \\
  \hline
 \end{tabular}
 \end{table}

\begin{figure}[ht]

 \includegraphics[width=7cm]{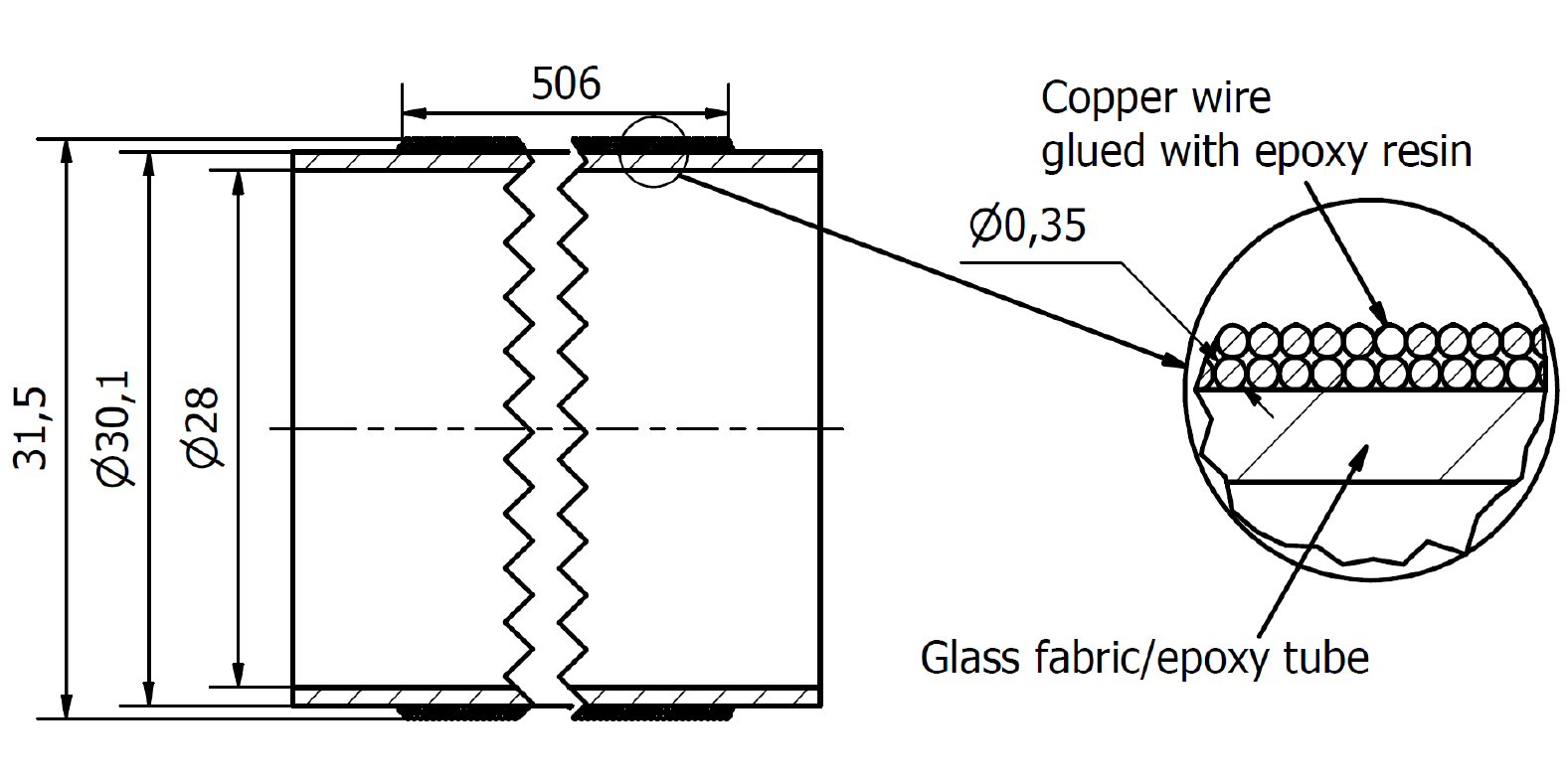}

  \caption{Sketch of the calibration solenoid.} \label{SO}
\end{figure}

The first layer of the calibration solenoid is winded on a glass fabric/epoxy tube and fixed with epoxy. The second layer, also fixed with epoxy is winded on the first layer after rectification of the additional fixation epoxy to obtain a diameter as regular as possible. The solenoid sketch is shown in Fig. \ref{SO}.

Using textbook formulas for a solenoid of finite length and taking into account the experimental error on the construction parameters, the field at the center of it on the symmetry axis is such that the ratio between the driving current and the obtained field R$_{B/I}$ is 7.253(3)~mT/A.

During calibration of the pick-up coil, the solenoid is driven by an alternating current of the order of 40~mA at frequencies varying in the range of several tens of Hz. The value of the driving current is measured with a commercial instrument whose accuracy is 0.06~$\%$  The signal at the ends of the pick up coil is demodulated using a lock-in amplifier. The accuracy of this instrument for voltage measurements is  0.2~$\%$. This is the limiting accuracy for the pick-up coil calibration. The measured product of the number of turns times the pick-up surface is 0.005215(10)~m$^2$. This value of the pick-up coil equivalent surface is used to recover the magnetic field value of the pulsed magnet, which is therefore given with respect to the one calculated for the calibration solenoid.

An important point is that we assume that in the calibration solenoid the radial homogeneity of the magnetic field is such that it can be neglected and we assume that the field is constant all over the pick-up surface. The homogeneity of the calibration solenoid is expected to be a fraction of a ppm.

The homogeneity of the 60~T pulsed magnet is such that a correction has to be considered when comparing the field given by the pick-up and the one given by the Rb sensor since the latter is only sensible to the magnetic field homogeneity in a scale shorter than a millimeter \textit{i.e.} about 20~ppm. In fact, to recover the magnetic field from the pick-up coil signal one assumes that the field is constant all over the pick-up surface, which is not exact for the 60 T pulsed magnet. The radial profile of the field is parabolic and the field is slightly higher at the border of the pick-up than at its center. With the assumption that the field is constant, the field value inferred thanks to the pick-up coil has been therefore evaluated to be about 0.1~$\%$ bigger than the one in its center to which the Rb sensor is sensible.

\section{Experimental Set-up}

\begin{figure}[ht]
\center
 \includegraphics[width=7cm]{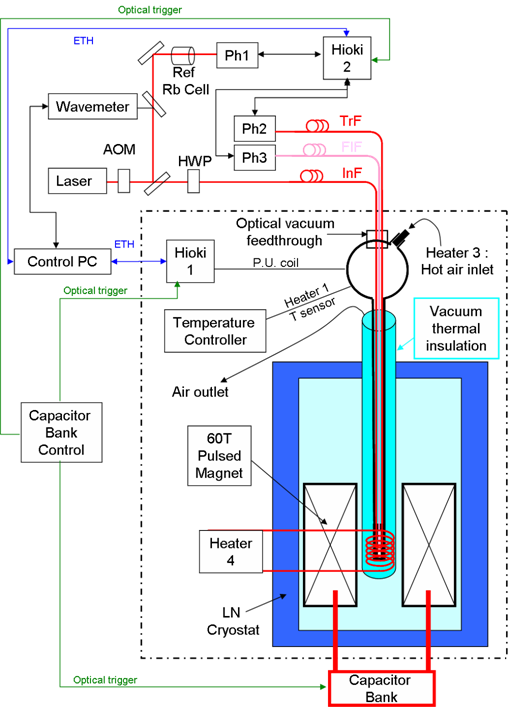}
  \caption{A view of the whole experimental set-up. On the top the optical elements with the laser light generation, its measurement by the wavemeter, the control of a reference Rb cell, and injection into the fiber transmitting light to the sensor. On the bottom the magnet components and the temperature regulations. All the instruments within the dot-dashed line  are inside a safety box }\label{SM}
\end{figure}

A view of the whole experimental set-up is in Fig.~\ref{SM}. The light beam coming from a DLX Toptica laser is sent to 1) a reference Rb cell contained within a mu-metal shield, 2)  a commercial wavemeter monitoring its wavelength continuously, 3)  a single-mode  fiber to transport it to the sensor after passing through a half wave plate  (HWP). The HWP rotation modifies the light polarization in order to control the light transmitted by the input polarizer shown in Fig.~3 top.  The transmission from the reference cell is detected by a photodiode (Ph1), while transmission of the sensor Rb cell and fluorescence are monitored by photodiodes Ph2 and Ph3. Ph1 and Ph2 are standard silicon photodiodes. Ph3 is a low noise variable gain photoreceiver with a -3 db optical electrical bandwidth of 7 Khz and a gain of $10^{10}$~V/W. All these signal are stored in a computer (Control PC) via a Hioki oscilloscope (Hioki 2) which also monitor the trigger signal given  by the Capacitor Bank Control delivering the optical trigger to start the magnetic pulse.

All the instruments  within the dashed lines of Fig.~\ref{SM} are actually inside a sealed box  not accessible during the magnet operation for safety reasons. The connection between the box and the outside takes place via another Hioki oscilloscope (Hioki 1) under control from the Control PC. The two Hioki oscilloscopes are synchronized at the microsecond level.\\
\indent Let's note that a heater on the internal side of the cryostat tail, indicated as Heater 4 in Fig. \ref{SM}, is used to obtain a better control of the whole probe temperature. It consists of a constantan wire wrapped around the inner tube in the insulating vacuum of the cryostat with a 3~W heating power.

\section{Results}

\begin{figure}[ht]
\includegraphics[width=8cm]{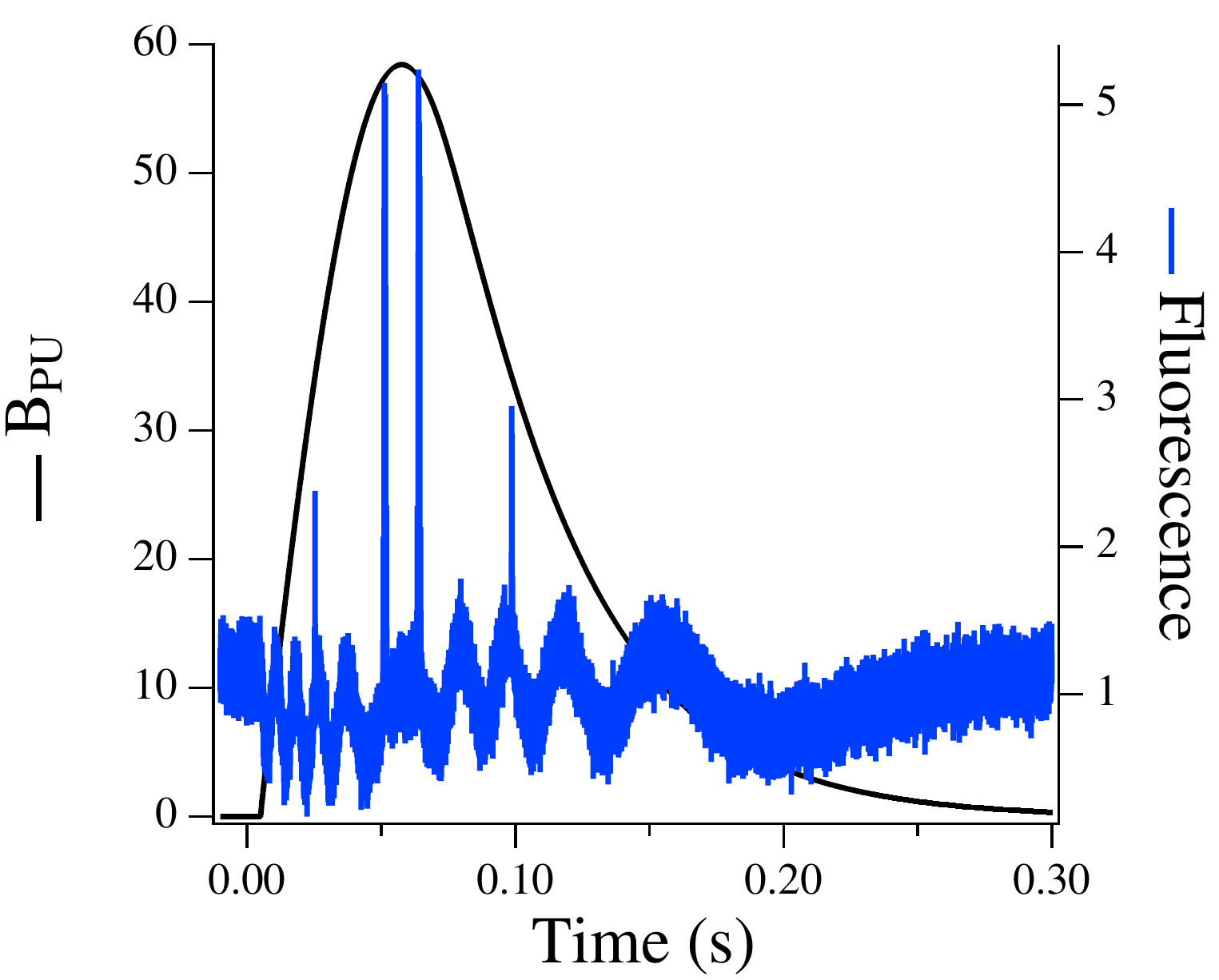}
 \includegraphics[width=8cm]{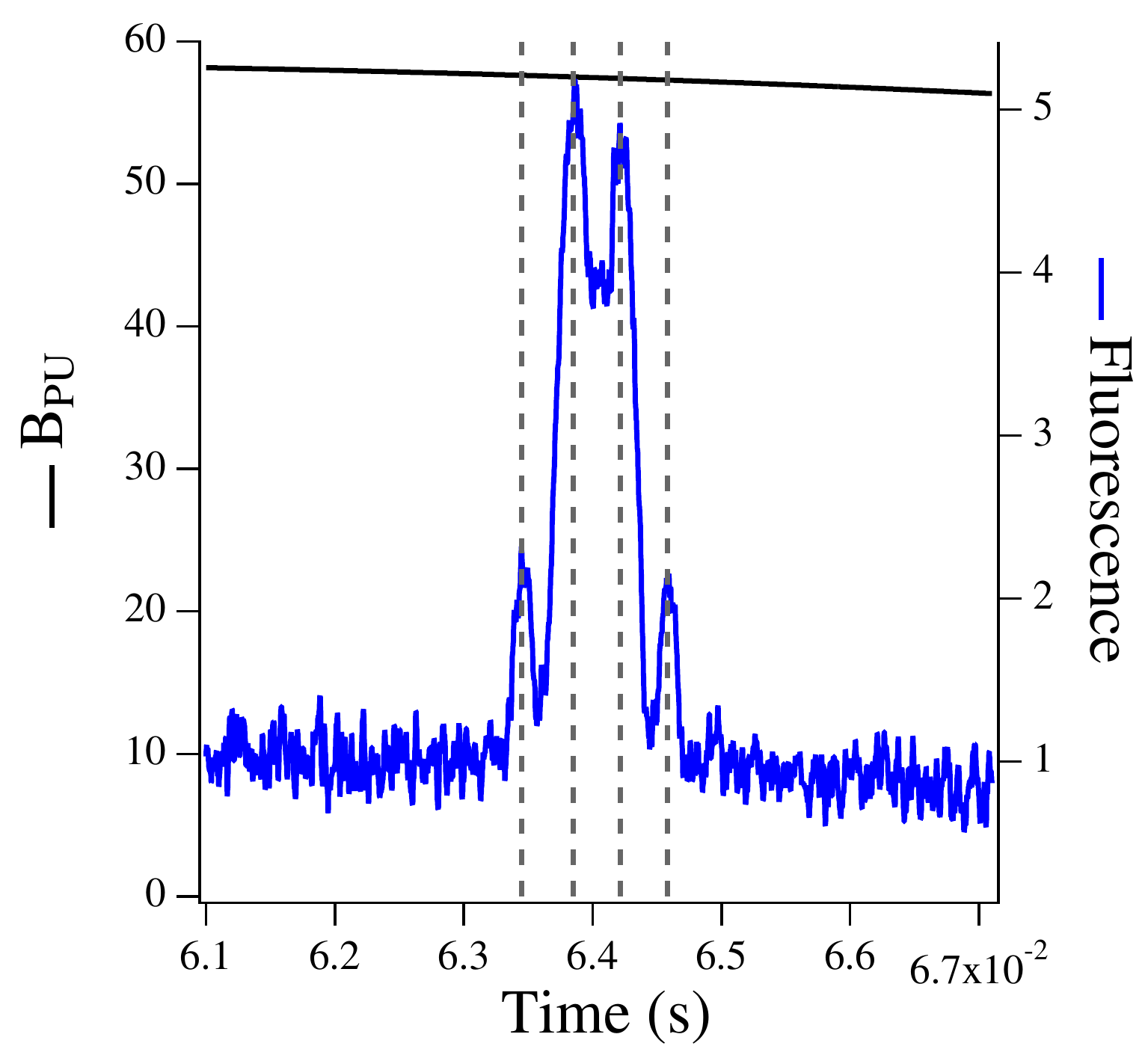}
\caption{On the top a typical rubidium fluorescence signal in Volt during the magnetic field pulse (in blue). The black line reports the $B_\mathrm{PU}$ field value in Tesla derived from the pick-up coil signal. On the bottom an expanded view of the signals presented on the top around one of the largest Rubidium absorptions, that  on the right at decreasing magnetic field.  The vertical light gray lines represent the center of each $^{87}$Rb resonance of Table~\ref{TableField}  determined by a Gaussian fit.  These data have been obtained under excitation by  a laser at frequency
$\nu=385042.737(5) \mathrm{GHz}$, where the uncertainty is given by the absolute accuracy of the wavelength meter.}
 \label{segnale}
\end{figure}

We recorded Rubidium fluorescence spectra and pick-up signals during several magnetic field pulses with different maximum fields. Figure \ref{segnale} shows a typical data acquisition, with the full temporal record on the top and and expanded view around the maximum magnetic field on the bottom. In black, the magnetic field strength derived by the pick-up signal is reported; it allows to record the temporal shape of magnetic field pulse.
\indent The blue traces in Fig.~\ref{segnale} shows the fluorescence signal. We observe four narrow Rubidium resonance peaks, two of them during the rise of the field and the other ones during the decreasing phase of the pulse. Each peak is composed by the superposition of four resonances, resolved in the expanded view on the bottom of that  figure. The resonances appearing at higher magnetic field correspond to the Rb $\sigma^+$ transitions $\vert J_g=1/2, m_g=1/2\rangle \rightarrow \vert J_e=3/2, m_e=3/2\rangle$, with a structure  produced by the Zeeman nuclear splitting.  The resonances observed at lower magnetic field correspond to the $\sigma^+$ transitions $\vert J_g=1/2, m_g=-1/2\rangle \rightarrow \vert J_e=3/2, m_e=1/2\rangle$ of both isotopes, with an unresolved nuclear spin structure on the figure scale. The atomic signal lies on an oscillating background due to the light on the glass cell surfaces.  The observed periodic oscillations are produced by Faraday effect on the light passing through the input single mode fiber and propagating parallel to the magnetic field direction. The periodic rotation of the light polarization leads to a periodic light transmission through the polarizer at the cell output. The oscillating signal is reduced by a proper orientation of the HWP at the single mode fibre input.

The bottom part of Fig.~\ref{segnale}  reports a zoom of the higher field data presented in the top part. We distinguish four resonances for rising and descending magnetic field. These resonances correspond to the $^{87}$Rb transitions listed in Table~\ref{TableField} listed by the increasing magnetic field strength. In addition, approximately at the center of the group of four $^{87}$Rb resonances, there are six resonance of $^{85}$Rb unresolved because their mutual separation is smaller than the Doppler width. Their presence produces an almost flat offset for the two central $^{87}$Rb resonances. In principle this fact could affect the position of the observed center of the involved resonances.

As pointed out previously, the transition between $\left|J_g=1/2, I=3/2,m_g=1/2, m_I=3/2\right\rangle$ and $\left|J_e=3/2, I=3/2;   m_e=3/2, m_I=3/2\right\rangle$  experiences a linear frequency shift for any value of the magnetic field. Following Eq.~\eqref{nu},
at the given laser frequency, the center of the considered resonance appears at the magnetic field reported on the bottom line of Table~\ref{TableField}. By  combining
this magnetic field calibration to the pick-up coil signal we derive the magnetic field position of all four high-field resonances, their  uncertainties derived from the combiation of Eqs.~\eqref{Errrand} and ~\eqref{Errsyst}.

\begin{table}%[H] add [H] placement to break table across pages
\caption{Quantum numbers and measured $B$ field position of the $^{87}$Rb high field resonances at $\Delta\nu_z$=812.331(5) GHz.}
\label{TableField}
\begin{ruledtabular}
\begin{tabular}{ccc}
Ground & Excited  & Magnetic \\
quantum numbers& quantum numbers & field\\
$|J_g,m_g;I,m_I\rangle$& $|J_e,m_e;I,m_I\rangle$ & T\\
$|1/2,1/2;3/2,-3/2\rangle$  &$|3/2,3/2;3/2,-3/2\rangle$  &57.865(13)\\
 $|1/2,1/2;3/2,-1/2\rangle$ &$|3/2,3/2;3/2,-1/2\rangle$  &57.979(13)\\
$|1/2,1/2;3/2,1/2\rangle$  &$|3/2,3/2;3/2,1/2\rangle$   &58.092(13) \\
$|1/2,1/2;3/2,3/2\rangle$   &$|3/2,3/2;3/2,3/2\rangle$   &58.204(13)\\
 \end{tabular}
 \end{ruledtabular}
 \end{table}

\begin{figure}
\begin{center}
 \includegraphics[width=7cm]{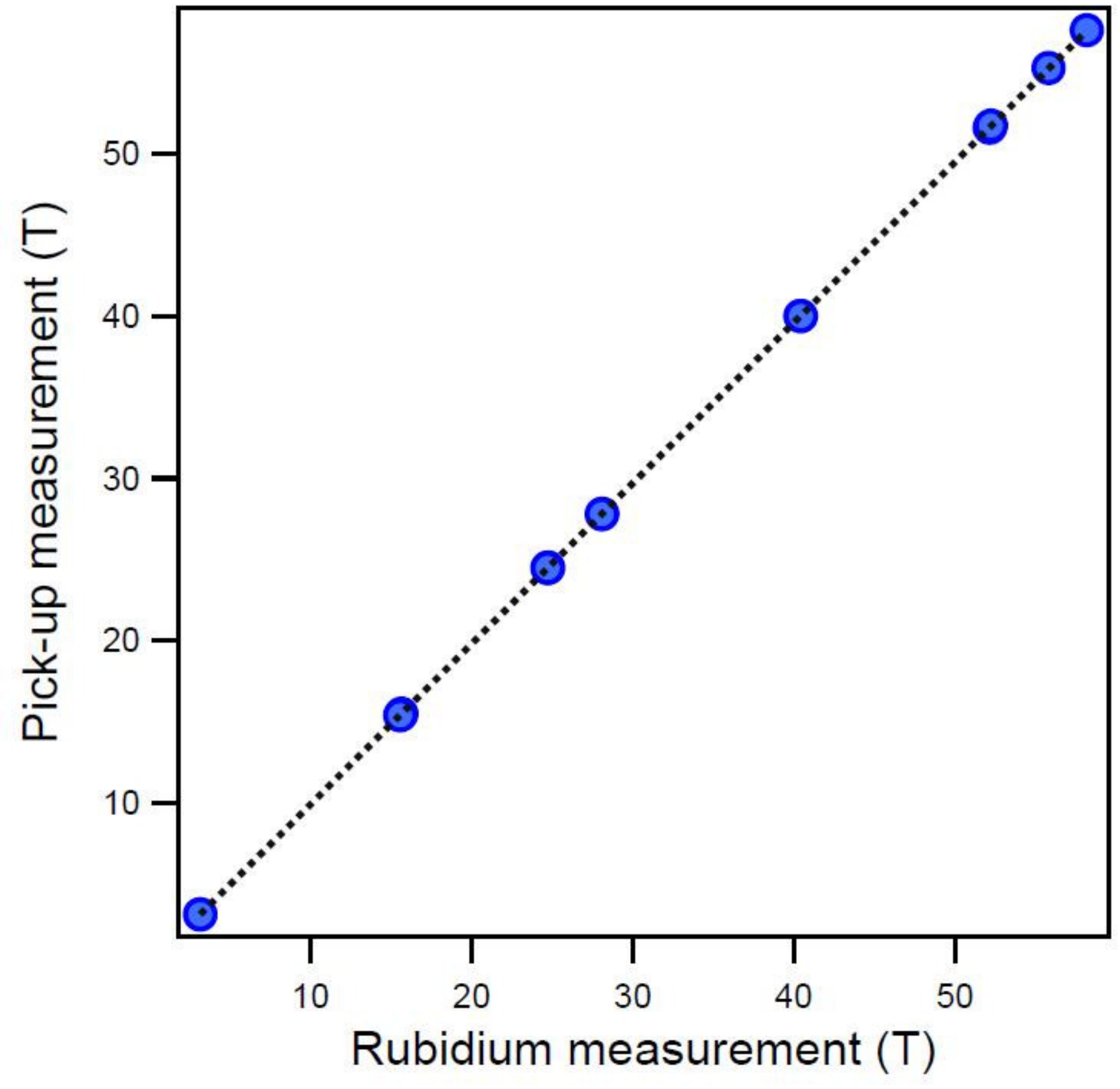}
  \end{center}
  \caption{Magnetic field strength measured by the pick-up coil as a function of the magnetic field strength given by Rubidium. Error bars on the horizontal axis, 200~ppm, and on the vertical axis, 0.22$\%$, are not shown because smaller than the markers. The linear fit is given by the equation reported within the text.}\label{PUvsRB}
\end{figure}

Using these data we scale the temporal profile of the magnetic field pulse given by the pick-up coil; thus we obtain the magnetic field strength during the whole duration of the pulse with an accuracy of about $2\times10^{-4}$, more than one order of magnitude better than the calibrated pick-up coil. This accuracy is limited by the knowledge of the $g_{5P}$ constant and not by our experimental method.

The value of the magnetic field given by our Rb sensor $B_{\mathrm{Rb}}$ has been compared to the one provided by the pick-up coil $B_{\mathrm{PU}}$ at several values of magnetic fields. The  $B_{\mathrm{PU}}$ values have been obtained by taking into consideration the effect due to the size of the pick-up explained in a previous section and the fact that the pick-up coil is located 7.5~mm above the probed atomic volume. Because of the pick-up position with respect to the Rb sensor the value given by the pick-up is 0.2$\%$ smaller than the Rb one as inferred from both the calculation of the longitudinal magnetic field profile of the coil and the
direct measurement performed moving the pick-up coil along the axis of the magnet.

Figure \ref{PUvsRB} reports the value of the magnetic field $B_{\mathrm{PU}}$ measured by the pick-up coil as a function of the values $B_{\mathrm{Rb}}$ given by Rubidium spectra. The linear relation is given by fitting the data taking into account the uncertainty of $B_{\mathrm{PU}}$ given by pick-up coil calibration and pick-up coil signal measurement. The uncertainty due to the pick-up calibration is 0.2$\%$ as explained before, the uncertainty of the measurement of the pick-up signal is about $0.1\%$, due to the voltage measurement. The final uncertainty is therefore 0.22$\%$.  Following the linear fit shown in fig.~\ref{PUvsRB}, the value of $B_{\mathrm{PU}}$ for $B_{\mathrm{Rb}}=0$ is compatible with zero within the error. The $B_{\mathrm{PU}}$ value can therefore be inferred by the $B_{\mathrm{Rb}}$ measurement thanks to the equation $B_{\mathrm{PU}} = 1.0009(6)B_{\mathrm{Rb}}$, which demonstrates the good agreement between  pick-up coil and our Rubidium sensor measurements.

\section{Conclusions}

Our Rb sensor shows uncertainties that are already more than an order of magnitudes better than the one of standard pick-up coils and also a direct access to a micrometer size explored region. It is  worth to stress that our experiment leads to a metrological measurement as far as accuracy is concerned. Indeed our measurement is a conversion between the Tesla unit and the frequency unit which can be related to the standard of time. This makes it a very interesting candidate to establish a secondary standard for the definition of the Tesla unit also taking into account that the accuracy of our system will improve in a straightforward way once a more accurate value of the g-factor of the excited states of the Rubidium will be available. In any case, our accuracy is much better than the one obtained in the measurement of high magnetic fields in the case of destructive fields that is limited to about 10$\%$.\\
\indent Our work may be compared with the recent work by Raithel research group \cite{Ma2017} where the sub-Doppler feature of electromagnetic induced transparency for two-photon transitions to Rydberg states were used for magnetic field measurements. The sub-Doppler spectroscopy leads to an increase resolution by a factor hundred. However the low accuracy of all the atomic constants associated to the Rydberg states cannot be compared to that for the ground and first-excited states. In our setup the observation of sub-Doppler absorption features rely on technical improvements realizable in the near future. Another straightforward way to improve our system is to move to a cell containing a single isotope of Rubidium to simplify the shape of the spectroscopy signal.\\
\indent As for the perspectives in physics measurements, we have succeeded in observing Rubidium transition at more than 58 T that is a world record as far non destructive field generation is concerned with an uncertainty per pulse never reached before.  This opens the way to dilute matter optical tests in high magnetic fields as precise measurements of g-factors of excited states at a level that is interesting to verify Quantum Electrodynamics predictions. In the way toward a sensor of a very high accuracy, which looks possible with our method, a measurement of the Rubidium diamagnetism will be necessary and our techniques combining atomic spectroscopy to non destructive pulsed high magnetic fields will certainly play a role also in this domain.

\section{Acknowledgments}
This research has been partially supported through NEXT (Grant No. ANR-10-LABX-0037) in the framework of the "Programme des Investissements d'Avenir". E.A. acknowledges a financial support from the "Chaires D'Excellence Pierre de Fermat" of the "Conseil Regional Midi-Pyren\'ees", France. S.S. acknowledges financial support from "Universit\'e Franco Italienne". The authors thanks M. Badalassi (INO-CNR, Pisa) and N. Puccini (Universit\'a di Pisa) for developing and testing  the mini-cell of the present investigation, F. Thibout for filling the cell with Rubidium at "Laboratoire Kastler-Brossel" (LKB, UPMC, Paris), G. Ballon and F. Giquel of the LNCMI-Toulouse for the help in the probe construction, and R. Mathevet for support. The PLA structure has been realized using a 3D printer at the CampusFab of the "Universit\'e Toulouse III".

\end{document}